\documentclass[aps,pra,twocolumn]{revtex4-2}

\usepackage[draft]{changes}
\definechangesauthor[name={Benedikt}, color=purple]{BT}
\definechangesauthor[name={Anders}, color=orange]{AS}

\usepackage[]{enumitem}
\usepackage[]{graphicx}
\usepackage[dvipsnames]{xcolor}

\usepackage[colorlinks=true,citecolor=NavyBlue,linkcolor=RubineRed,urlcolor=Cerulean,pdfencoding=auto]{hyperref}
\usepackage[]{amsmath}
\usepackage[]{amssymb}

\def\bra#1{\mathinner{\langle{#1}|}}
\def\ket#1{\mathinner{|{#1}\rangle}}

\def\proj#1{\ket{#1}\bra{#1}}
\def\op#1#2{\ket{#1}\bra{#2}}

\let\Re\undefined

\DeclareMathOperator{\Re}{Re}

\begin{document}
\title{Quantum Networks with Realistic Photonic Memories} 

\author{Benedikt Tissot}

\author{Sophia Carmen Simon}

\author{Anders Søndberg Sørensen}
\email[]{anders.sorensen@nbi.ku.dk}
\affiliation{Center for Hybrid Quantum Networks (Hy-Q), Niels Bohr Institute, University of Copenhagen, Jagtvej 155A, DK-2200 Copenhagen, Denmark}

\begin{abstract}
  Ensemble based photonic memories paired with photon pair sources are the essential ingredient for many quantum networking protocols.
  Most protocols and theoretical analyses focus on networks of identical nodes, however,
  realistic networks consist of nodes that may vary from device to device.
  We investigate a network of non-identical memories and develop mitigation strategies for the variation in performance.
  We find that by strategically distributing the memories among the nodes and individually adapting the emission probabilities based on the memory efficiency the effect of the variance on fidelity and duration can to a large extend be mitigated.
  Generally, our analysis shows the importance of accounting and compensating for device variations to avoid deteriorating the performance.
  Lastly, we also link our results to memories where efficiency decays as a function of storage duration.
  Here, we show that our approach provides a visible improvement for networks with non-identical and decaying memories.
\end{abstract}
\maketitle

\section{Introduction}

Quantum networks or the even more advanced quantum internet promise access to secure communication \cite{ekert-1991-quant_crypt_based_bells_theor,acin-2007-devic_indep_secur_quant_crypt,gisin-2007-quant_commun,pirandola-2020-advan_quant_crypt}, distributed quantum computing \cite{cuomo-2020-towar_distr_quant_comput_ecosy}, and enhanced sensing and metrology \cite{komar-2014-quant_networ_clock,brady-2026-correl_noise_estim_with_quant_sensor_networ,li-2026-multip_estim_with_array_entan_atomic_sensor,stas-2026-entan_assis_non_local_optic} by providing distant coherently linked quantum systems as a resource.
Additionally, these networks can make the observation of Bell inequality violations accessible over large distances  \cite{duan-2001-long_distan_quant_commun_with,clauser-1969-propos_exper_to_test_local}.
A central building block for long-distance quantum networking is the ability to generate, store, and retrieve entanglement between remote nodes with high rates and fidelity.
Entanglement generation has been achieved in a growing list of systems, e.g., 
trapped ions   \cite{moehring-2007-entan_singl_atom_quant_bits_at_distan,stephenson-2020-high_rate_high_fidel_entan,krutyanskiy-2023-entan_trapp_ion_qubit_separ_by_meter},
atomic ensembles \cite{chou-2007-funct_quant_nodes_entan_distr,yuan-2008-exper_demon_bdcz_quant_repeat_node,yu-2020-entan_two_quant_memor_via}, 
color centers in solids \cite{bernien-2013-heral_entan_between_solid_state,sipahigil-2016-integ_diamon_nanop_platf_quant_optic_networ,humphreys-2018-deter_deliv_remot_entan_quant_networ}, and rare earth ions \cite{lago-rivera-2021-telec_heral_entan_between_multim,liu-2021-heral_entan_distr_between_two,ruskuc-2025-multip_entan_multi_emitt_quant_networ_nodes}.
For concreteness, we will focus on extensions of the seminal proposal by Duan, Lukin, Cirac, and Zoller (DLCZ) \cite{duan-2001-long_distan_quant_commun_with}, which introduced a practical architecture for quantum repeaters based on atomic ensembles, linear optics, and single-photon detection, enabling entanglement distribution through heralded photon-mediated schemes.
Several improvements, e.g., in entanglement generation \cite{zhao-2007-robus_creat_entan_between_remot_memor_qubit} or swapping \cite{jiang-2007-fast_robus_approac_to_long} or by using photonic memories \cite{simon-2007-quant_repeat_with_photon_pair,sangouard-2008-robus_effic_quant_repeat_with} (see Ref.~\cite{sangouard-2011-quant_repeat_based_atomic_ensem_linear_optic} for a review), have been proposed.
More recently, these approaches have been investigated in combination with single atomic systems \cite{gu-2025-hybrid_quant_repeat_with_ensem,cussenot-2025-unitin_quant_proces_nodes_cavit,tissot-2025-hybrid_singl_ion_atomic_ensem,tissot-2025-singl_doubl_click_high_rate,sun-2025-hybrid_quant_repeat_chain_with}.

One of the adaptations of the original DLCZ protocol is to replace the atomic ensembles with spontaneous parametric down conversion sources (SPDC) and (rare-earth ion doped crystal) quantum memories (QM) \cite{simon-2007-quant_repeat_with_photon_pair}.
Central to the benefit of this proposal is that the SPDC+QM nodes are compatible with telecommunication wavelength, support temporal, spectral, and spatial multiplexing as well as high duty cycles \cite{simon-2007-quant_repeat_with_photon_pair,sinclair-2014-spect_multip_scalab_quant_photon,lago-rivera-2021-telec_heral_entan_between_multim,liu-2021-heral_entan_distr_between_two,ruskuc-2025-multip_entan_multi_emitt_quant_networ_nodes}, 
which can dramatically enhance entanglement distribution rates.
Furthermore, SPDC+QM have demonstrated long storage durations \cite{ruskuc-2022-nuclear_spin_wave_quant_regis,zhong-2015-optic_addres_nuclear_spins_solid}.
The combination of QMs capable of storing photonic qubits with SPDC sources provide a promising route toward modular and heterogeneous quantum networks.

In theoretical studies of imperfect systems, the parameters are typically taken to be identical for all components (see Ref.~\cite{sangouard-2011-quant_repeat_based_atomic_ensem_linear_optic} for a review for quantum repeaters).
In practice, however, different QMs will e.g. have different efficiencies.

To the best of our knowledge the potential impact of non-identical memories has remained largely unexplored.
A recent work in this direction investigated decoherence aware routing in quantum networks \cite{huang-2026-decoh_aware_entan_swapp_strat}.
In this article we investigate non-uniform DLCZ-type quantum networks with non-identical SPDC+QM nodes including the effect on the fundamental entanglement generation step.
The nodes may differ in their storage efficiency as well as photon pair generation rate.
Such non-uniformity breaks the symmetry assumptions often adopted in idealized repeater analyses and introduces new trade-offs in entanglement generation rates, fidelity, and resource allocation.
We develop a theoretical framework to model heralded entanglement generation between non-identical remote SPDC+QM nodes, explicitly accounting for higher-order pair emissions, non-uniform efficiencies, and time-dependent memory decay.
Based on our analysis, we propose mitigation strategies compensating for non-uniform efficiencies, both by strategic distribution of the memories in the network nodes and adapting the individual emission rates of the SPDC's of the different nodes.

The remaining article first introduces the full protocol (Sec.~\ref{sec:protocol}),
which is modeled in Sec.~\ref{sec:model}.
Using our model we first provide a leading order analytic analysis (Sec.~\ref{sec:analysis}) which we use in Sec.~\ref{sec:mitigation} to develop mitigation strategies for networks with device-to-device varying memory efficiencies.
In Sec.~\ref{sec:results} we use this to simulate a concrete network and test our mitigation strategies leading to a conclusion in Sec.~\ref{sec:conclusion}.

\section{Protocol}\label{sec:protocol}
\begin{figure}
\centering

\includegraphics[width=\linewidth]{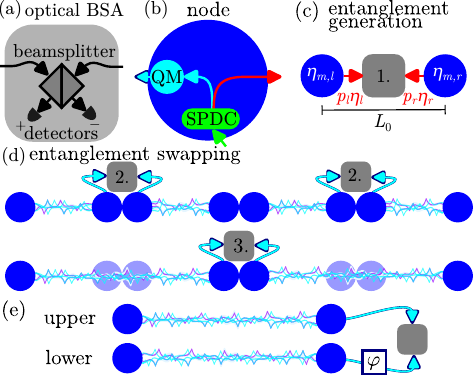}
\caption{\label{fig:protocol-overview} Overview of the DLCZ protocol implemented using multi-mode quantum memories.
  The building blocks for the protocol are (a) probabilistic optical BSAs, implemented by combining two optical channels on a beamsplitter terminating in detectors \(\pm\), and (b) the nodes made up by an SPDC photon pair source and a QM.
  The first step of the protocol is entanglement generation (c), where two nodes each generate a photon pair. Here we consider that the left \(l\) and right \(r\) node can have different emission probabilities \(p_{s}\) with \(s=l,r\).
  Furthermore, we account for differing memory (transmission) efficiencies \(\eta_{m,s}\) (\(\eta_s\)) of the nodes (to the heralding detection).
  Reading out the memories enables (d) entanglement swapping to extend the range of a single rail.
  Each rail is split into multiple (\(2^n\) for swap level \(n\)) fundamental links, and iteratively swapped to the edges of the network. Here we illustrate swap level \(n=2\), where 4 entangled (shown as wiggly lines) fundamental links per rail are first pairwise swapped (2.) and a final swap in the center (3.) leads to edge-to-edge entanglement.
  (e) In the DLCZ protocol two rails are generated in parallel, resulting in a dual rail setup.
  A node of the upper and lower rail are combined, and the state with a photon in the upper (lower) rail and vacuum in the other rail encodes the dual-rail qubit states \(\ket{1}\) (\(\ket{0}\)), which can, e.g., be read out by varying the phase $\varphi$ in one arm before combining them on a beam splitter.
}
\end{figure}

Before we delve into the detailed modeling and analysis of non-uniform quantum networks, we first introduce the particular protocol we will investigate.
This protocol is a memory based implementation of the DLCZ protocol \cite{duan-2001-long_distan_quant_commun_with},
where SPDCs and multimode QMs are used to increase the achievable rate \cite{simon-2007-quant_repeat_with_photon_pair}.
The multiplexing enhancement can be understood as changing the rate of attempts of a single qubit (or mode) from \(c/L\) over a distance \(L\) with \(c\) being the speed of light 
to \(N c / L\) for \(N\) multiplexed qubits (or modes) \cite{simon-2007-quant_repeat_with_photon_pair,ruskuc-2025-multip_entan_multi_emitt_quant_networ_nodes}.
For such setups, the influence of the multi-modal nature of the SPDC+QM nodes was recently investigated in Ref.~\cite{hellebek-2025-impac_multim_sourc_dlcz_type_quant_repeat}.
In the following we will build upon these insights and introduce the protocol and our model,
where we incorporate the possibility of differing memory efficiencies.

The central building blocks of the protocol, are the SPDC+QM nodes as well as optical Bell state analysers (BSA), see Fig.~\ref{fig:protocol-overview}(a) and (b).
The latter is made up by a 50:50 optical beamsplitter (BS) combining two inputs and terminating in two detectors, which we label \(\pm\).
A single click detection removes the which path information of a single photon arriving from the left or right and can therefore be used to generate or swap entanglement.
In the following we will assume photon-number resolving detectors, noting 
that this can be achieved with state of the art (multi-pixel) superconducting nanowire single-photon detectors \cite{esmaeil2021superconducting,gruenenfelder-2023-fast_singl_photon_detec_real}.
Using these building blocks the protocol distributes entanglement in a dual rail setup between SPDC+QM nodes with SPDC+QM nodes also acting as repeaters, see Fig.~\ref{fig:protocol-overview}(c).
In this repeater network, entanglement is first generated between \emph{fundamental links}, and then pairwise swapped to the outer nodes of neighboring links.
To exploit parallelization in generation and swapping, each swap level halves the complete network, see Fig.~\ref{fig:protocol-overview}(d).
Considering that a network generates entanglement over a distance \(L\),
a swap level of \(n\) reduces the distance for entanglement generation in the fundamental links to \(L_0 = L/2^n\) 
and requires \(2^n+1\) repeater stations and \(2 \cdot 2 \cdot 2^n\) SPDC+QM nodes for a dual rail setup.

\section{Model}\label{sec:model}

In the following we will detail the steps of the protocol outlined in the previous section and the mathematical model used to describe them.
In the absence of loss we describe the nodes with a model following single-mode statistics, i.e., states of the form
\begin{align}
  \label{eq:node-state}
\left[ \sqrt{1 - p - p^2} \ket{0} + \sqrt{p} a^{\dag} \ket{1} + p a^{\dag} a^{\dag} / \sqrt{2} \ket{2} \right] \ket{\emptyset} ,
\end{align}
where \(\ket{k}\) denotes \(k=0,1,2\) photons stored in the memory of the node and the \(a^{\dag}\) denotes the creation of a photon travelling to the heralding station.
The probability to emit a single photon is given by \(p\), note that this is an effective parameter.
In general sources may not be single mode.
In Ref.~\cite{hellebek-2024-charac_multim_natur_singl_photon} it was, however, shown that despite their inherent multimode nature a single mode description can be justified even for commonly used continuously driven SPDCs sources provided that we chose a suitable effective number of modes for the source.
We will therefore employ this single mode description.

\subsection{Fundamental Link}\label{sec:generation}

Having introduced the model for the ideal nodes (before photon transmission), we now turn to the model for the entanglement generation in the fundamental links.
For now, we assume ideal storage in the memory, but we will introduce a limited memory efficiency below.
We allow for different emission probabilities \(p_l\) and \(p_r\) for the left and right node (see Fig.~\ref{fig:protocol-overview}) as well as different efficiencies for the photons used for entanglement generation.
The heralding is performed by combining the channels while removing the which path information on a 50:50 BS followed by detectors with annihilation operators \(d_{\pm}\).
The map from the initial left channel can thus be written as
a transformation of the corresponding photon annihilation operator 
\(a \to \sqrt{\eta_l/2} (d_+ + d_-) + \sqrt{1-\eta_l} a_L\) and analogous for the annihilation operator for right channel \(b \to \sqrt{\eta_r/2} (d_+ - d_-) + \sqrt{1-\eta_r} b_L\), where we include all loss processes affecting the heralding photon in the efficiencies \(\eta_s\) (\(s=l,r\)) with corresponding loss operators \(a_L\) and \(b_L\).
We take the corresponding bosonic reservoirs to be in the vacuum state.
With this we calculate the non-normalized state of the fundamental node assuming photon-number resolving detection and phase stability of the difference in length of the two distances, resulting in
\begin{widetext}
\begin{align}
  \label{eq:fundamental-state}
\frac{P_f}{2} \rho_f = 
  & \frac{1}{2} \left\{ [\sqrt{\eta_l p_l (1 - p_r)} \ket{1,0} \pm \sqrt{\eta_r p_r (1 - p_l)} \ket{0,1}] \cdot [\text{H.c.}] \right\}
    + \frac{1}{2} (1-\eta_l) \Big[ \left( \sqrt{\eta_r p_l p_r} \ket{1,1} \pm \sqrt{2 \eta_l p_l^2} \ket{2,0} \right) \cdot (\text{H.c.}) \Big]  \notag \\
  & + \frac{1}{2} (1-\eta_r) \Big[ \left( \sqrt{\eta_l p_l p_r} \ket{1,1} \pm \sqrt{2 \eta_r p_r^2} \ket{0,2} \right) \cdot (\text{H.c.}) \Big] 
    + p_{d,f} \proj{0,0} 
    + \text{higher orders} .
\end{align}
\end{widetext}
Here, the states \(\ket{k,q}\) denote the Fock states of \(k=0,1,2\) (\(q=0,1,2\)) photons stored in the memory to the left (right) of the link.
We have here introduced a perturbative treatment of dark counts with the probability to have a dark count in one detector within the click window of a single attempt \(p_{d,f}\) being treated as a perturbative parameter.
In the expression we write out terms up to second order in emission probabilities and first order in dark counts.
We will use this approximate expression as the basis for the following analysis.
While one might consider assymetric placement of the nodes and thus differing transmission efficiencies (\(\eta_l\) and \(\eta_r\)) to compensate for differing memories,
such assymetries increase the entanglement duration
 \cite{avis-2024-asymm_node_placem_fiber_based_quant_networ},
highlighting why we allow for different emission probabilities instead.
We note that 
in realistic networks different transmission efficiencies  might be unavoidable.

Before we discuss the memory losses and the remaining steps of the full protocol in more detail, we note that up to the above perturbative order, the density matrix can be written in the form
\begin{align}
  \label{eq:state}
  \rho = & \sum_{k,q} D_{k,q} \proj{k,q} + \left( C \op{0,1}{1,0} + \text{H.c.} \right)\\
  & + \left( c \op{2,0}{1,1} + c' \op{0,2}{1,1} + d \op{2,1}{1,2} + \text{H.c.} \right) , \notag
\end{align}
with coherences (off-diagonal components) \(C, c, c', d\) and diagonal entries \(D_{k,q}\) with \(k,q=0,1,2\).
Within our perturbative treatment only elements \(k+q \le 2\) have a contribution.
Note that, assuming phase stability we can choose $C$ to be real within the model.
In an ideal, uniform network, the fundamental links prepare the state \(D_{1,0} = D_{0,1} = C = 1/2\), and the remaining elements are \(0\).
Some of those elements are still \(0\) in the state generated in the fundamental link \eqref{eq:fundamental-state} but can become non-zero due to memory losses and during entanglement swapping as discussed in the following.

\subsection{Memory Losses}\label{sec:memory-loss}
For ensemble based memories loss is the main decoherence mechanism \cite{zhao-2008-millis_quant_memor_scalab_quant_networ,albrecht-2015-contr_rephas_singl_collec_spin}
and we will therefore focus on it in the following.
The memory efficiencies are taken into account using a linear optics model,
where we include intrinsic inefficiencies of the memory and the decay during storage.
We emphasize that this accounts for the different memory efficiencies of the individual memories as well as any dependence of the retrieval efficiency on the storage time.

For the memory with label \(s=l,r\) we model the losses analogous to the heralding channel, denoting the creation operator of the memory as \(c_s^\dag \to \sqrt{\eta_{m,s}} c_{s}^\dag + \sqrt{1-\eta_{m,s}} c_{s,L}^\dag\)  which is linked to the basis states via \(\ket{k,q} = {( c_l^k c_r^q )^\dag} \ket{\emptyset}/{\sqrt{k! q!}}\) and \(c_{s,L}\) denote the annihilation operators for the loss channels.
The efficiency \(\eta_{m,s}\) unites the efficiency of all processes from collection up to detection.
Since the memories are far apart, we treat the left and right memory losses as independent processes and 
for a state of the form of Eq.~\eqref{eq:state} both of those processes result in a state of the same form.
Therefore, we formulate both as an update rule for the state.

To account for the efficiency of the left memory of a link, we find the update rule for the matrix elements.
The matrix elements are mapped from the state before the decay (right hand side below) to the decayed state (left hand side below),
according to the update rule:
\begin{align}
  \label{eq:left-memory-decay}
  \begin{aligned}
    D_{k,q} & \leftarrow \sum_{\tilde{k} \ge k} \frac{\tilde{k}!}{(\tilde{k}-k)! k!} \eta_{m,l}^k (1-\eta_{m,l})^{\tilde{k}-k} D_{\tilde{k},q} , \\
    C & \leftarrow \eta_{m,l}^{1/2} [C + \sqrt{2} (1-\eta_{m,l}) c^{*}] , \\
    c' & \leftarrow \eta_{m,l}^{1/2} [c' + \sqrt{2} (1-\eta_{m,l}) d^{*}] , \\
    c & \leftarrow \eta_{m,l}^{3/2} c ,
    \qquad \qquad d \leftarrow \eta_{m,l}^{3/2} d .
  \end{aligned}
\end{align}

Analogously, accounting for the efficiency of the right memory of a link leads to the update rule
\begin{align}
  \label{eq:right-memory-loss}
  \begin{aligned}
    D_{k,q} & \leftarrow \sum_{\tilde{q} \ge q} \frac{\tilde{q}!}{(\tilde{q}-q)! q!} \eta_{m,r}^q (1-\eta_{m,r})^{\tilde{q}-q} D_{k,\tilde{q}} , \\
    C & \leftarrow \eta_{m,r}^{1/2} [C + \sqrt{2} (1-\eta_{m,r}) c'] , \\
    c & \leftarrow \eta_{m,r}^{1/2} [c + \sqrt{2} (1-\eta_{m,r}) d] , \\
    c' & \leftarrow \eta_{m,r}^{3/2} c', 
    \qquad \qquad d \leftarrow \eta_{m,r}^{3/2} d .
  \end{aligned}
\end{align}

For both left and right memory loss channels we see, that the higher photon number subspaces decay to all lower photon subspaces.
This includes re-population of the target coherence \(C\) from higher lying coherences \(c\) or \(c'\), as well as the repopulation of $c$ and $c'$ from \(d\).
However, this re-population occurs with an amplitude \(\sqrt{2 \eta_{m,s}} (1-\eta_{m,s}) < 1\) for \(\eta_{m,s} \in (0,1]\) and \(s=l,r\).
The maximum is \((2/3)^{3/2} < 1\) for \(\eta_{m,s} = 1/3\).
Although this repopulation of the coherence slightly mitigate the effect of loss on the coherence,
this shows that (not surprisingly) increasing the SPDC emission probability cannot arbitrarily counteract photon losses.

\subsection{Entanglement Swapping}\label{sec:swap}
In order to swap the entanglement to the outer edges of two neighboring fundamental links, the memories are read out and combined on a 50:50 BS terminating in detectors.
Depending on which detector clicked, a difference in phase is introduced, see Eqs.~\eqref{eq:fundamental-state} and \eqref{eq:state}.
Assuming that this phase is either tracked or corrected, we continue the calculation only with the state without the phase and account for the other possibility by doubling the success probability.
Considering two states of the form in Eq.~\eqref{eq:state} being merged by  entanglement swapping, results in an extended state \(\rho_S\) of the same form.
We thus also express the swap in terms of an update rule for the non-normalized density matrix \(P_S \rho_S/2\) and denote quantities of one of the links using a tilde above them \(\tilde{\cdot}\).
Using the trace of the non-normalized density matrix we can calculate the success probability \(P_S\) and the normalized density matrix.

The update rule merging the two links (right hand side of the expression below) into an extended link (left hand side of the expression below) for photon-number-resolving detection is then
\begin{align}
  \label{eq:entanglement-swapping}
  \begin{aligned}
\frac{P_S}{2} D_{k,q} & \leftarrow p_{d,m} D_{k,0} \tilde{D}_{0,q} + \frac{1}{2} \sum_{w=0,1} D_{k,w} \tilde{D}_{1-w,q} , \\
\frac{P_S}{2} C & \leftarrow \frac{1}{2} C \tilde{C} , 
\qquad \qquad \frac{P_S}{2} c \leftarrow \frac{1}{2} c_L {\tilde{C}}^{*} , \\
\frac{P_S}{2} c' & \leftarrow \frac{1}{2} C_L \tilde{c}' , 
\qquad \qquad \frac{P_S}{2} d \leftarrow \frac{1}{2} c_L \tilde{c}^{\prime *} . 
  \end{aligned}
\end{align}
Note that the different time scales of entanglement generation and memory read out but also different wavelengths and thus the associated differences in detector performance can lead to different probabilities of having a dark count in one detector during a memory photon detection \(p_{d,m}\) compared to during generation of the fundamental link \(p_{d,f}\). We have therefore introduced two different dark count probabilities for these processes. 

\subsection{Dual rail one photon per side subspace}\label{sec:dual-rail}
The entanglement generated in a state of the form of Eq.~\eqref{eq:state} is photon-number-encoded single-rail entanglement, which is experimentally hard to process in a non-local setting.
One way for processing single-rail entanglement is transfering to atomic systems \cite{cussenot-2025-unitin_quant_proces_nodes_cavit,tissot-2025-hybrid_singl_ion_atomic_ensem,tissot-2025-singl_doubl_click_high_rate,sun-2025-hybrid_quant_repeat_chain_with}.
A simpler possibility, however, is to use a dual rail setup as proposed in the seminal DLCZ protocol \cite{duan-2001-long_distan_quant_commun_with}.
These approaches enable state readout in different bases, which we show below, can be used to, e.g., observe Bell inequality violations.
The dual rail setup created by two states of the form of Eq.~\eqref{eq:state} in a parallel geometry [see Fig.~\ref{fig:protocol-overview}(e)]
is post selected such that each end has exactly one excitation shared between the two rails.
Ideally, this corresponds to combining two rails such that the resulting combined state is described by
\begin{align}
  \label{eq:state-dual-rail}
  \rho = & \! \sum_{k,q=0,1} \hspace{-2mm} D_{k,q} \proj{k,q}_d + \left( C \op{0,1}{1,0}_d + \text{H.c.} \right) ,
\end{align}
i.e., of the form of Eq.~\eqref{eq:state} but with several elements vanishing.
Therefore, we can express the state using another update rule
(where the ``lower'' rail is denoted with a tilde above it \(\tilde{\cdot}\))
\begin{align}
  \label{eq:post-selection}
  \begin{aligned}
  P_P D_{k,q} & \leftarrow D_{k,q} \tilde{D}_{1-k,1-q} \text{ for } k,q=0,1 ; \\
  P_P C & \leftarrow C \tilde{C}^{*} .
  \end{aligned}
\end{align}
Here, the basis states now denote logical dual-rail qubits \(\ket{k,q}_d = \ket{k, \widetilde{1-k}, q, \widetilde{1-q}}\),
i.e., the logical qubit state \(\ket{0}_d\) (\(\ket{1}_d\)) has the photon stored in the ``lower'' (``upper'') rail.
While the single photon per side state (and thus the density matrix) is not directly accessible,
it can be accessed by post selection (where it is then also affected by dark-counts).
The dark-count contribution to the single-photon per side density matrix can be modeled by adding \(D^d_{k,q} = p_{d,m} [D_{0,q} \tilde{D}_{0,1-q} + D_{k,0} \tilde{D}_{1-k,0}] + p_{d,m}^2 D_{0,0} \tilde{D}_{0,0}\) to the non-normalised density matrix.
The resulting post-selected density matrix models experiments using photon-number resolving detectors, in accordance to the previous steps.

In addition, to the probability \(P_P\) for the state to be in the post-selected subspace,
another figure of merit is the (post-selected) fidelity
\begin{align}
  \label{eq:post-selected-fidelity}
  F = \frac{D_{1,0} + D_{0,1}}{2}  + \Re C .
\end{align}

We will focus on the fidelity \eqref{eq:post-selected-fidelity} as an easily accessible quantity for our analysis and optimization.

\subsection{Full network description}
In the previous sections we have discussed how to model the entanglement generation within non-uniform fundamental links, the effect of (non-uniform) finite memory efficiencies, entanglement swapping, as well as reading out and post selecting the state in the single photon subspace when employing a dual rail architecture.
These can be combined to study a full network of swap depth \(n\) by first using the generation model for all individual fundamental links, which results in a set of generation success probabilities \(P_k\) and states \(\rho_k\) for the link \(k=1,\dots,2^{n+1}\).
The probabilities \(P_k\) determine the geometric distribution describing the probability that the \(k\)'th link successfully generates entanglement in the \(q\)th trial \((1-P_k)^q P_k\) resulting in a state modelled by the density matrix \(\rho_k\).

Successfully generating entanglement at two links that are supposed to be merged according to the current swap level (see Fig.~\ref{fig:protocol-overview}) is followed by the probabilistic entanglement swapping which succeeds with probability \(P_S\) according to Eq.~\eqref{eq:entanglement-swapping}.
Note that we leave the dependence on the links implicit in denoting \(P_S\), but the different entanglement swaps can occur with different success probabilities, due to the differences in the links.
Furthermore, to account for memory losses during swapping, the states are updated to account for the finite efficiency of the memories read out during the entanglement swap.
We emphasize that this enables inserting the different memory efficiencies as time-dependent functions to account for memory decay.
Following this, the update rules modelling the swap can be applied to calculate the success probability and the heralded state after swapping.
To model the full network this process is repeated until the final swap level is reached.
Finally, after the dual rail state preparation is accounted for analogously to the swaps, we can quantify the state using the single-photon-per-side post selection probability and fidelity.

In the following analysis we explore two approaches to implement all these step in a numerical simulation.
If the memories can be described with a time-independent efficiency, we use a Markov chain model \cite{shchukin-2019-waitin_time_quant_repeat_with} to calculate the average duration to prepare the final state. 
This enables us to calculate the average duration and fidelity without any excess noise due to random sampling.

If, however, the memory efficiencies are time-dependent, not only do the swap probabilities depend on the storage duration but also the final fidelity depends on the pattern of click times of the various steps.
In this case, we implement a numeric Monte Carlo simulation, averaging over individual trajectories to estimate the average duration and fidelity.
We highlight, that while the Markov chain approach enables calculating fidelity and average duration without noise, this approach becomes much more complicated for time-dependent memories.
Therefore we resorted to Monte Carlo simulation in many situations for simplicity. 
Note that the Monte Carlo approach can also have performance benefits for low success probabilities (high target fidelities), where the Markov chain becomes increasingly long.
However, the noise introduced by Monte Carlo sampling to calculate averages,
makes numerical optimization harder.
Therefore, the better suited method depends both on the necessary noise level,
as well as the parameter regime.
This underlines the value of having both approaches available.

\section{Analytic Analysis}\label{sec:analysis}

Using the modeling framework discussed above, we will now delve into the detailed analysis of a concrete network.
As we aim to use both numerical and analytical calculations to gain understanding, we will focus on swap depth \(n=1\), resulting in a network made up by 8 SPDC+QM nodes as show in Fig.~\ref{fig:eight-node-network}.

\subsection{Uniform Efficiencies}
We begin our analysis with the simple case of uniform memories.
Uniform optical quantum networks have been studied extensively in the literature, see Ref.~\cite{sangouard-2011-quant_repeat_based_atomic_ensem_linear_optic} for a review,
therefore we only emphasize the key aspects here.
First, for the single-click protocol under consideration for entanglement generation and in the absence of dark-counts we find that the leading order success probability of the fundamental link generation is given by \(P_f \approx 2 \eta \bar{p}\) if all nodes share the same emission probability \(\bar{p}\) and link efficiency \(\eta\).
Furthermore, the density matrix elements of the fundamental link including the memory efficiencies are given by
\begin{align}
  \label{eq:uniform-efficiencies-FL}
  \begin{aligned}
  D_{0,0} & = \frac{1-\bar{\eta}_m}{2} [1 - 3(1-\eta) \bar{p}] + 3 (1-\eta) (1-\bar{\eta}_m)^2 \bar{p} , \hspace{-5mm}\\
  D_{0,1} & 
            = \frac{\bar{\eta}_m}{2} [1 - 3(1-\eta) \bar{p}] + 3 (1-\eta) (1-\bar{\eta}_m) \bar{\eta}_m \bar{p} , \\
  C & = \frac{\bar{\eta}_m}{2} [1 - 3(1-\eta) \bar{p}] + 2 (1-\eta) (1-\bar{\eta}_m) \bar{\eta}_m \bar{p} , \\
  D_{1,1} & 
            = (1-\eta) \bar{\eta}_m^2 \bar{p} ,
  \qquad \qquad c  
      = \frac{\bar{\eta}_m^2}{\sqrt{2}} \left(1 - \eta \right) \bar{p} ,
  \end{aligned}
\end{align}
and \(c' = c\), \(D_{2,0} = D_{0,2} = D_{1,1}\).
With this density matrix
and accounting for the above update rules for swapping and post-selection
we calculate the final infidelity of the uniform eight node network
\begin{align}
  \label{eq:uniform-infidelity}
  1-F_{\text{uniform}} = 24 (1-\bar{\eta}_m) (1-\eta) \bar{p} ,
\end{align}
to first order in the emission probability \(\bar{p}\).
The infidelity increases if the memory efficiency \(\bar{\eta}_m\) decreases and decreases with increasing emission probability \(\bar{p}\).
The latter leads to a trade-off between rate and fidelity. 
In the following we will use \(\bar{x}\) to denote the average quantities of an instance.

\subsection{Non-uniform Efficiencies}

We expect small errors to be linearly additive, such that in this regime the
(homogeneously) calculated error, proportional to the emission probability \(\bar{p}\),
can be combined with other errors, e.g., the leading order error introduced by the device-to-device variance calculated in the following.
Calculating the leading error introduced by variance in the network, enables devising mitigation strategies to achieve high fidelities and improved rate fidelity trade-offs.
For the current perturbative treatment we again disregard dark counts and
consider \(\bar{p} \to 0\) while fixing \(p_r/p_l\), such that we are able to calculate the intrinsic error due to a non-uniform network.

To investigate this regime we first calculate the leading order of the density matrix elements [see Eq.~\eqref{eq:state}] of a fundamental link.
The non-vanishing elements in the leading order are
\begin{align}
  \label{eq:non-uniform-FL}
  \begin{aligned}
D_{0,0} & = 1 - \frac{\eta_{m,l} p_l + \eta_{m,r} p_r}{p_l + p_r} ,
 \qquad D_{1,0} = \frac{\eta_{m,l} p_l}{p_l+p_r} , \\
D_{0,1} & = \frac{\eta_{m,r} p_r}{p_l + p_r} ,
 \qquad \qquad \qquad C = \frac{\sqrt{\eta_{m,l} \eta_{m,r} p_l p_r}}{p_l+p_r}. 
  \end{aligned}
\end{align}
Note that the leading order success probability of the link generation is linear in the average emission probability \(\bar{p}\) or equivalently the individual emission probabilities \(P_f = \eta (p_l + p_r)\)

with the efficiency from emission to heralding detection \(\eta\) being equal on each side.
As noted above differing efficiencies of the photons used for the fundamental generation is detrimental, such that we here choose equal efficiencies for the heralding photons \(\eta_l = \eta_r = \eta\) corresponding to e.g. identical distances to the heralding stations. The effect of different heralding efficiencies, introduced by having different distances was considered in \cite{avis-2024-asymm_node_placem_fiber_based_quant_networ}. Here we introduce an imbalance via the emission probabilities to counteract the device-to-device variance in memory efficiencies and hence we only consider the effect of differences between $p_l$ and $p_r$ as well as \(\eta_{m,l}\) and \(\eta_{m,r}\).

The leading order entanglement swap success probability (\(n=1\)) is
\begin{align}
  \label{eq:non-uniform-swap-probability}
  P_S = \frac{\eta_{m,r} p_r}{p_l + p_r} + \frac{\tilde{\eta}_{m,l} \tilde{p}_l}{\tilde{p}_l + \tilde{p}_r} - 2 \frac{\eta_{m,r} p_r}{p_l + p_r} \frac{\tilde{\eta}_{m,l} \tilde{p}_l}{\tilde{p}_l + \tilde{p}_r} ,
\end{align}
where we denote the fundamental links as above and mark one of them using a tilde \(\tilde{\cdot}\) [as in Eq.~\eqref{eq:entanglement-swapping}].
The matrix elements and swap success probability after entanglement swapping (\(n=1\)) are
\begin{align}
C & = \frac{\sqrt{\eta_{m,l} \eta_{m,r} \tilde{\eta}_{m,l} \tilde{\eta}_{m,r} p_l p_r \tilde{p}_l \tilde{p}_r}}{\tilde{\eta}_{m,l}(p_l+p_r) \tilde{p}_l + \eta_{m,r} p_r (\tilde{p}_l+\tilde{p}_r) - 2 \eta_{m,r} \tilde{\eta}_{m,l} p_r \tilde{p}_l} , \notag \\
D_{0,1} & = \frac{\eta_{m,r} \tilde{\eta}_{m,r} p_r \tilde{p}_r}{\tilde{\eta}_{m,l}(p_l+p_r) \tilde{p}_l + \eta_{m,r} p_r (\tilde{p}_l+\tilde{p}_r) - 2 \eta_{m,r} \tilde{\eta}_{m,l} p_r \tilde{p}_l} , \notag \\
D_{1,0} & = \frac{\eta_{m,l} \tilde{\eta}_{m,l} p_l \tilde{p}_l}{\tilde{\eta}_{m,l}(p_l+p_r) \tilde{p}_l + \eta_{m,r} p_r (\tilde{p}_l+\tilde{p}_r) - 2 \eta_{m,r} \tilde{\eta}_{m,l} p_r \tilde{p}_l} , \notag \\
  \label{eq:non-uniform-swapped}
D_{0,0} & = 1 - D_{0,1} - D_{1,0} .
\end{align}

\begin{figure}
\centering

\includegraphics[width=\linewidth]{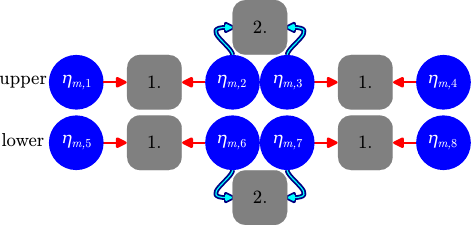}
\caption{\label{fig:eight-node-network} Illustration of an eight node dual rail network.
  In an eight node network entanglement is generated pairwise between neighboring nodes (1.).
  Upon successfully generating these links within a rail entanglement is swapped (2.) to the outer nodes.
  This sketch uses the building blocks of the DLCZ protocol introduced in Fig.~\ref{fig:protocol-overview}.}
\end{figure}
Using the above calculation for a single rail, we can express the final state of a 8-node dual-rail network in terms of the efficiencies of all memories.
To this end we enumerate the memory efficiencies \(\eta_{m,q}\) and emission probabilities \(p_q\) for \(q=1,\dots,8\) from top left to bottom right (see Fig.~\ref{fig:eight-node-network}).
The success probabilities for entanglement generation of the \(k=1,2,3,4\) fundamental link are then
\begin{align}
  \label{eq:Pgen_leading}
  P_k = \eta (p_{2k-1} + p_{2k}) ,
\end{align}
and the swap probabilities in rail \(R=1,2\)
\begin{align}
  \label{eq:Pswap_leading}
  P_{S,R} = & \frac{\eta_{m, 4R-2} p_{4R-2}}{p_{4R-3} + p_{4R-2}} + \frac{\eta_{m,4R-1} p_{4R-1}}{p_{4R-1} + p_{4R}}
              \notag \\ &
                          - 2 \frac{\eta_{m, 4R-2} p_{4R-2}}{p_{4R-3} + p_{4R-2}} \frac{\eta_{m,4R-1} p_{4R-1}}{p_{4R-1} + p_{4R}} .
\end{align}
Finally, the probability for the final state to have one photon per side is
\begin{align}
  \label{eq:Ppost_leading}
  P_P =  \frac{\left( \prod_{i=1,3,6,8} p_i \eta_{m,i} \right) + \left( \prod_{i=2,4,5,7} p_i \eta_{m,i} \right)}{P_{S1} P_{S2} \prod_{k=1}^4 (p_{2k-1}+p_{2k})} .
\end{align}
The leading order fidelity of the generated state (in the subspace with a single photons at each end) upon succeeding all steps is
\begin{align}
  \label{eq:dual_rail_leading_fidelity}
  F = \frac{1}{2} + \frac{\sqrt{\prod_{i=1}^8 p_i \eta_{m,i}}}{\left( \underset{i=1,3,6,8}{\prod} p_i \eta_{m,i} \right) + \left( \underset{i=2,4,5,7}{\prod} p_i \eta_{m,i} \right)} .
\end{align}
We emphasize that if the memory efficiencies \(\eta_{m,q}\) are storage duration dependent, all of the above quantities apart from the generation probability, become trajectory dependent with the efficiency evaluated according to the click time of the corresponding photons.

\section{Mitigation approaches}\label{sec:mitigation}
To develop mitigation strategies, we consider the case, where the \(\eta_{m,q}\) are time-independent, but are non-uniform.
Physically this corresponds to the scenario where there is some variance in device production, but the coherence time largely exceeds the average time of entanglement generation.
In this scenario we can investigate several different approaches to mitigate the effect of varying memory efficiencies.
We consider two control nobs: the placement of nodes in the network and choosing non-uniform emission probabilities.
For the node placement we keep the distance between nodes fixed, but compare two approaches:
\begin{enumerate}
  \item The \emph{naive} approach of placing the memories randomly within the network.
 \label{list:nosort}
  \item Placing the memories \emph{sorted} in the network guided by our leading order fidelity expression [Eq.~\eqref{eq:dual_rail_leading_fidelity}],
    i.e., targeting \(\prod_{i=1,3,6,8} \eta_i = \prod_{i=2,4,5,7} \eta_i\).
    For concreteness we choose a sorting of the form \(\eta_8 \ge \eta_4 \ge \eta_5 \ge \eta_1 \ge \eta_6 \ge \eta_2 \ge \eta_7 \ge \eta_3\).
 \label{list:sort}
\end{enumerate}
Although, neither of those may be the ideal choice, for a particular random choice of $\{\eta_i\}$ it gives a simple strategy that allows us to understand the rough performance with non-uniform efficiencies.
We note that in addition to using the fidelity as a guide, we also incorporate the expectation that due to parallelization, 
 the average total duration is less affected by memory inefficiencies of  the swap than by the inefficiency of post selection.
This can be seen by using the approach in Ref.~\cite{avis-2024-asymm_node_placem_fiber_based_quant_networ} to estimate the average duration to prepare two different tasks in parallel.
Given that the upper (lower) rail finishes at a rate $\mathbf{R}_1$ ($\mathbf{R}_2$) and that up to leading order $\mathbf{R}_R \propto P_{S,R}$, preparing both simultaneously leads to the rate for both to be ready
$(\mathbf{R}_1 + \mathbf{R}_2) \left( 1 + \frac{\mathbf{R}_1}{\mathbf{R}_2} + \frac{\mathbf{R}_2}{\mathbf{R}_1} \right)^{-1}$, which is linear in the memory efficiencies (in the leading order)  since \(R_l \propto \eta_i\).
In contrast the post selection probability scales quadratically in the memory efficiencies.
Intuitively, this can be understood by post-selection requiring two simultaneous clicks, while the swaps are decoupled by the parallelization and each require only a single click.
Therefore, we put the more efficient memories at the edge of the network.
Considering that the sorting approach can also be understood in terms of pairwise minimizing the variance between the rails and also minimizing the asymmetry between the two (ideally mirror symmetric) halves merged by the final swap, we expect a similar approach to be readily applicable to higher swap levels.

In addition, to the memory sorting we also investigate four approaches to choose (non-uniform) emission probabilities:
\begin{enumerate}[label=\alph*.]
  \item The \emph{naive} approach is to choose the emission probabilities uniform \(p_i = p\).
 \label{list:uniform}
  \item The \emph{symmetric links} approach, where we choose the emission probabilities to lead to equal read-out photon probability between nodes of the same links, i.e., \(\frac{p_{2k-1}}{p_{2k}} = \frac{\eta_{m,2k}}{\eta_{m,2k-1}}\). This approach fully mitigates the leading order infidelity due to the non-uniform efficiencies, i.e. it leads to \(F=1\) in Eq.~\eqref{eq:dual_rail_leading_fidelity}.
 \label{list:symmetric}
  \item The \emph{biased links} approach, where we 
    
    bias the links towards efficient memories by choosing
    \(\frac{p_{2k-1}}{p_{2k}} = \frac{\eta_{m,2k-1}}{\eta_{m,2k}}\).
 \label{list:biased}
  \item Lastly, a full \emph{numeric} optimization of the emission probabilities to minimize the duration it takes to achieve a link with a certain target fidelity.
 \label{list:numeric}
\end{enumerate}
While in principle all emission approaches can be used for both naive and sorted placement of memories,
the biased links approach is primarily intended for sorted placement.
The heuristic approaches \ref{list:uniform}, \ref{list:symmetric}, and \ref{list:biased} are designed to use local link information such that they can readily be embedded in more complex network architectures.

\section{Results}\label{sec:results}
\subsection{Variations from device to device}
\begin{figure}
\centering
\includegraphics[width=\linewidth]{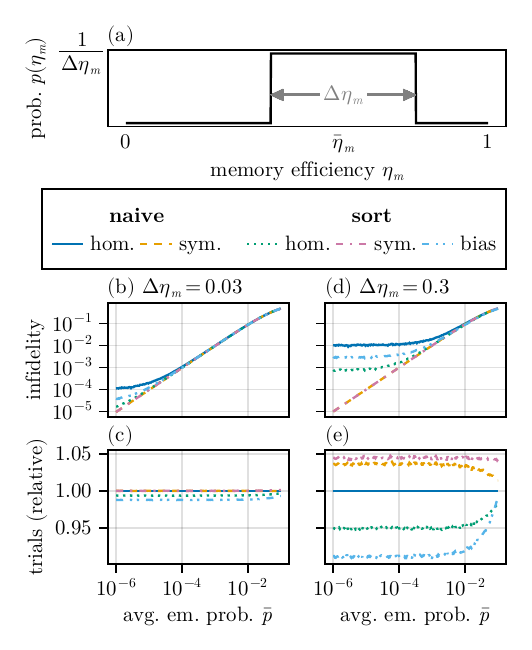}
\caption{\label{fig:emission-prob-infidelity} 
(a) Considered probability distribution  for the memory efficiency and (b-e)
average infidelity and relative trials until success as a function of emission probability for randomly drawn memory efficiencies with distribution mean $\bar{\eta}_m$ and width $\Delta \eta_m$.
The average is taken over \(1 000\) samples of 8 efficiencies drawn from a flat distribution with width \(\Delta \eta_m = 0.03\) for
(b) and (c) and \(\Delta \eta_m = 0.3\) for (d) and (e).
Panels (b), (d) show the infidelity as a function of the emission probability,  whereas (c), (e) compares the average time required for entanglement generation by considering
the average number of trials required for entanglement generation relative to the average trials of the naive placement with homogeneous emission probability.
}
\end{figure}
To understand the effect of non-uniform memories and to benchmark the mitigation approaches we will first investigate the average fidelity for memory efficiencies drawn from a flat distribution [see Fig.~\ref{fig:emission-prob-infidelity} (a)] as a function of emission probability, see Fig.~\ref{fig:emission-prob-infidelity}(b) and (d).
Note, that this means that each of the calculated means has contributions from instances with different mean and variance of the memory efficiencies.
In all figures we assume the transmission efficiency \(\eta = e^{-L_0/(2L_{\text{att}})}\) and trial time \(T_T = L_0 / c\) to be dominated by the long distance fiber links of length \(L_0\) of the fundamental links.
We additionally use the speed of light in fiber \(c=2c_0/3\) (with \(c_0\) the speed of light in vacuum) and the fiber attenuation \(0.2\,\)dB\(/\)km corresponding to an attenuation length \(L_{\text{att}} \sim 22\,\)km. 
Furthermore, we assume a multiplexing capacity \(N = 1000\),
and to focus on the errors imposed by the memories, we consider no dark counts \(p_{d,f} = p_{d,m} = 0\).
For Fig.~\ref{fig:emission-prob-infidelity} 
we draw \(1000\) instances of memory efficiencies
and consider the fundamental links to span a distance \(L_0=200\,\)km (resulting in a full network distance of \(400\,\)km),
and the average of the memory efficiency distribution \(\bar{\eta}_m = 0.6\).

We observe in Fig.~\ref{fig:emission-prob-infidelity}(b,d)  that all approaches converge together for large probabilities, which we attribute to the two-photon error [see Eq.~\eqref{eq:uniform-infidelity}] dominating over the error introduced by the non-uniform network in this regime.
Additionally, we see that the curves for the different approaches tend to flatten out for low emission probabilities at different infidelities.
We can understand this with our leading order expression \eqref{eq:dual_rail_leading_fidelity}, which underlines that the approaches lead to different leading order infidelities.
By design the approaches that symmetrize the local links lead to \(F=1\) to the leading order,
which agrees with the corresponding curves retaining a linear dependence for low emission probabilities, while the other approaches flatten out at a finite infidelity.
  On the one hand, in the low fidelity regime (high emission probability) the infidelities of the different approaches are very similar, such that the percent level difference in duration becomes the deciding factor for the ideal protocol.
  On the other hand, not all approaches can reach high fidelities, such that if high fidelity is required the difference in trial times becomes insignificant.
When focusing only on the infidelity, the symmetric links approach \ref{list:symmetric} thus seems favorable.

\begin{figure*}
\centering
\includegraphics[width=\linewidth]{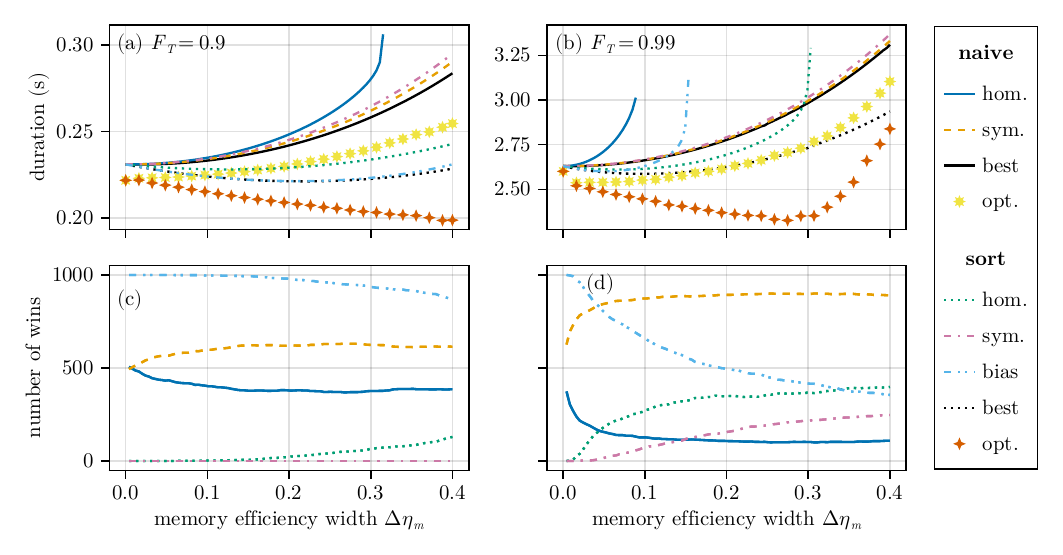}
\caption{\label{fig:opt-duration} (a,b) Average duration and (c,d) number of instances of the best manual optimization approach as a function of the width of the memory efficiency distribution.
  The average is taken over 1000 instances of 8 memory efficiencies of the network from a flat distribution around \(\bar{\eta}_m = 0.6\) with a varying variance as indicated on the \(x\)-axis.
  On the left (a,c) we target a fidelity \(F_T=0.9\) while on the right (b,d) we target a fidelity \(F_T=0.99\).
  We compare the different mitigation approaches (see legend), where the black lines correspond to the best heuristic approach, and the numeric approach for random memory placement takes the symmetrized approach while for the sorted placement it takes the biased (symmetrized) approach as an initial guess for \(F_T = 0.9\) (0.99).
  In panel (c,d)  we compare the number of instances where each approach performes best (wins) for the heuristic optimizations (approaches \ref{list:uniform}, \ref{list:symmetric}, and \ref{list:biased}).
  Note that we do this comparison individually for sorted placement and random placement of the memories so that the total count is 2000.
}
\end{figure*}
A possible way to compare the approaches that accounts for both the different success probabilities as well as the average infidelity of the protocol simultaneously,
is to target a fidelity \(F_T\) and optimize the emission probabilities to minimize the average duration required to achieve this fidelity.
To optimize the average duration when using the heuristic approaches \ref{list:uniform}, \ref{list:symmetric}, and \ref{list:biased} we use an approximate Newton method \cite{press-2007-numerical_recipes}
  to find the largest mean emission probability \(\bar{p}\) that achieves the target infidelity.
  This is possible because in the heuristic approaches the relation between different emission probabilities is directly fixed by the memory efficiencies.
For the numeric optimization \ref{list:numeric} we use the ``NelderMead()'' method implemented in ``Optim.jl'' \cite{mogensen-2018-julia_optim} in the Julia programming language \cite{bezanson-2017-julia}
to optimize all emission probabilities
with a cost function that allows violating the infidelity and trusted parameter regions at an (exponential) cost, i.e., a direct optimization encoding the constraints with (exponential) penalties \cite{fletcher-2013-practical_method_optimization}.
For more details we refer to Appendix \ref{app:opt}, and highlight here that the approximate Newton method approach only relies on calculating the fidelity, which for the time-independent efficiencies is possible without using the numeric evaluation of the Markov chain or Monte Carlo simulation.
This relies on the assumption that the fidelity and average duration are monotonous which the leading order expression fulfills for low infidelities.
Consequently, the approximate Newton method is computationally favorable in particular compared to the full numeric optimization.
For the full numerical optimization we use the result of one of our optimized mitigation approaches as the initial guess:
Concretely, for the naive approach of memory placement we base the optimization on the symmetrizing of the emission probabilties \ref{list:symmetric},
while for the sorted placement of memories we chose the biased approach \ref{list:biased} for the lower fidelity target ($F_T = 0.9$) and the symmetrizing approach \ref{list:symmetric} for the higher fidelity target ($F_T = 0.99$).

In Fig.~\ref{fig:opt-duration}(a) and (b) we display the average duration as a function of the width of the memory efficiency distribution.
Here we again average over 1000 instances of the 8 memory efficiencies drawn from a flat distribution around \(\bar{\eta}_m = 0.6\),
and we use the same parameters as for Fig.~\ref{fig:emission-prob-infidelity} but additionally target a fidelity of \(F_T=0.9\) (a) or \(F_T=0.99\) (b).
We see that the performance of the approaches aligns with the insights gained in Fig.~\ref{fig:emission-prob-infidelity}.
The best and fully numeric approaches can exploit the higher memory efficiencies obtained for some of the memories to obtain a better performance by sorting the memories.
At a certain point, however, the lowest memory efficiencies begin to limit the performance and the durations begin to sharply rise.
Naturally, the full numerical optimization on average leads to a performance increase compared to the heuristic approaches \ref{list:uniform}, \ref{list:symmetric}, and \ref{list:biased}.
Therefore, a full numerical optimization is worthwhile for the particular memories available in the lab.
For this reason we have made our code available at Ref.~\cite{CODE}, so that this optimization can easily be performed.

Since it can be hard to find the  global optimum,
the heuristic approaches serve as important starting points for the numerical optimization because they terminate deterministically.
However, we note that -- as visible by the sharp rise after \(\Delta \eta_m \sim 0.3\) for the higher target fidelity \(F_T = 0.99\) -- not all numerical optimizations converged (in which case we fall back to the initial guess).
Fig.~\ref{fig:opt-duration}(c) and (d) furthermore show that which of the heuristic approaches is better depends on the target fidelity as well as the particular instance of the memory efficiencies. 
Here we plot for how many instances an approach for choosing the emission probabilities outperforms the other approaches (i.e., ``wins'') for the same memory placement.
From these plots we see that there is no simple answer to the question of which approach is better, although the biased approach tends to be better at low fidelity and narrow distributions, whereas the symmetrized and homogenous perform better for broader distributions. 

Comparing the two target fidelities we also see that the sorted, biased approach that is predominantly favorable for the target fidelity \(F_T = 0.9\), diverges in average duration for the higher target fidelity \(F_T=0.99\).
This  (as well as the one of the naive, homogeneous approach for both fidelities and the sorted homogeneous approach for the higher fidelity) is caused by the protocol failing to achieve the target fidelity for some instances if the memory efficiency distribution becomes too wide.
This underlines the interplay between fidelity and success probabilities which needs to be balanced by the optimal approach.

In Fig.~\ref{fig:emission-prob-infidelity} and \ref{fig:opt-duration} we display data drawn from a fixed distribution with a certain average and width.
This means that we actually compare instances with different mean. 
In Appendix \ref{app:binning} we display the average duration based on the data of Fig.~\ref{fig:opt-duration} binned according to the average memory efficiency and the difference between the largest and lowest memory efficiency of the individual instances (instead of the underlying distribution we draw from).
This largely confirms the above insights, including the difference in how the approaches are affected by the spread of the memory efficiencies.

To summarize, for static memory efficiencies, our leading order analysis provide valuable insight to develop mitigation strategies for the variation in memory efficiency.
The naive approach of random placement and uniform emission probabilities often failed to reach the target fidelity,
while other strategies can still achieve the target fidelity.
This highlights that mitigation strategies may be indispensable for achieving the desired performance in real-world networks.
In particular, we have shown that both the placement of the memories as well as device-to-device varying emission probabilities provide a valuable control nob in realistic quantum networks.

\subsection{Time dependent efficiencies}
\begin{figure*}
\centering
\includegraphics[width=\linewidth]{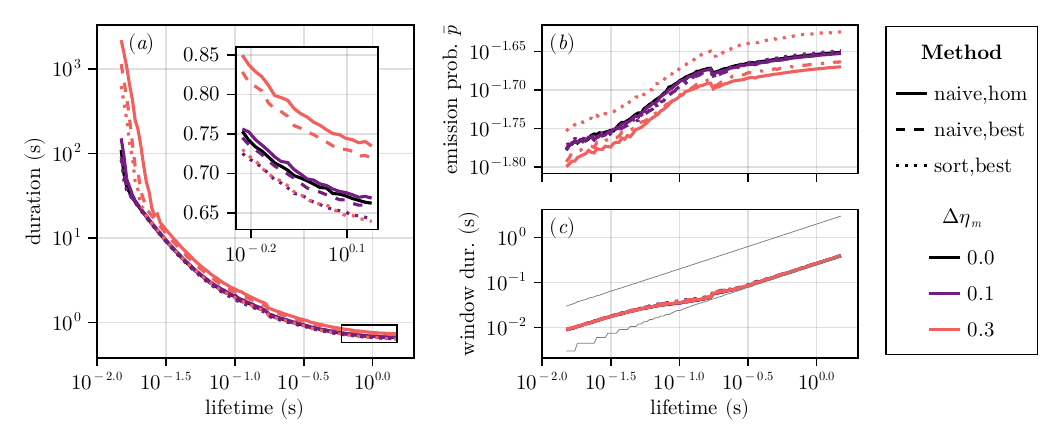}
\caption{\label{fig:opt-duration-decay} The effect of a finite memory lifetime on (a) the duration, (b) the average emission probability, and (c) the acceptance window averaged over \(200\) randomly drawn memory instances from a flat distribution with width \(\Delta \eta_m\) encoded by the color (see legend).
  We compare no mitigation (naive,hom.: solid line), to the best (heuristic) mitigation approaches for instances with sorted (dotted line) and naive memory placement (dashed line).
  The inset of panel (a) displays a zoom for longer lifetimes.
  In general the duration decreases for longer lifetimes, and the mitigation approaches are beneficial both for short and long lifetimes.
  The optimal acceptance window is roughly proportional to the lifetimes, and the average emission probability begins to saturate for long lifetimes.
  The gray line in panel (c) indicates the minimal (which is also the starting value of the optimization) and maximal window duration considered in the optimization.
}
\end{figure*}

We now turn to including the effect of time-dependent decay of the efficiency in our model.
This will decrease the average efficiency and can also introduce additional variance in the memory efficiencies, as the storage durations of the memories vary from node to node and trajectory to trajectory.
The latter also implies that all quantities apart from the generation success probability of the fundamental links, become dependent on the particular click times (or rather the time between relevant clicks), i.e., the individual trajectories.
Due to the trajectory dependence, we will use the Monte Carlo trajectory sampling method introduced in Sec.~\ref{sec:model}.
Not only does this make the calculation of the average fidelity numerically more expensive, but also adds noise due to the finite sampling of the fidelity and duration.
For this reason the full numerical optimization becomes infeasible
and we thus restrict ourselves to the heuristic approaches \ref{list:uniform}, \ref{list:symmetric}, and \ref{list:biased}.

To counteract the effect of the decaying efficiency we introduce an additional mitigation
step in the form of a temporal cut-off \cite{collins-2007-multip_memor_insen_quant_repeat,li-2021-effic_optim_cutof_quant_repeat_chain}, where links including memories that exceed a certain storage duration are reset and regenerated.
The cut-off duration (in units of trial-time) is an additional (discrete) optimization parameter.

To achieve a numerically feasible method, we assume that increasing the cut-off window always reduces the average fidelity.
We justify this by the fact, that for a cut-off window of a single trial, the decay does not introduce an additional variance and thus no leading order infidelity contribution.
Longer cut-offs increase the variance and decrease the average efficiency, both of which negatively affect the fidelity of the final state, see Eq.~\eqref{eq:uniform-infidelity} and \eqref{eq:dual_rail_leading_fidelity}.
Based on these observations we combine the approximate Newton method, used in the time-independent case, with an outer loop increasing the window.
If increasing the window size does not lead to a decreased duration (while achieving the target fidelity) two times in a row, we treat this as the optimal window duration.
We choose to compare with two steps not only one, to suppress noise from the Monte Carlo sampling.

For concreteness we assume that the temporal decay takes a Gaussian shape  \cite{zhao-2008-millis_quant_memor_scalab_quant_networ}, i.e., the time-dependent part of the memory efficiencies is \(e^{-(t/\tau)^2}\) with storage duration \(t\) and the lifetime \(\tau\).
Apart from introducing the lifetime, compared to the previous simulations
we reduce the drawn memory instances to \(200\), increase the memory efficiency distribution mean to \(\bar{\eta}_m = 0.8\) and the link distance to \(L_0 = 300\,\)km.
Additionally, we target a fidelity of \(F=0.9\), and create averages over \(500\) Monte Carlo trajectories.
To optimize the acceptance window we use up to \(200\) samples between \(\max \left( 1, \left\lfloor \frac{\tau}{4 T_T} \right\rfloor \right)\) and \(\left\lceil 2\frac{\tau}{T_T} \right\rceil\) with \(T_T\) the trial time.

We combine the memory decay with the device variance, by setting the total memory efficiency to be the product of the time dependent factor and an intrinsic efficiency drawn from a flat distribution analogous to the previous section.
In Fig.~\ref{fig:opt-duration-decay} we compare sorting and non-sorting, as well as homogeneous emission and the average of the best heuristic approach, for two different widths $\Delta \eta_m$ and,
for reference, also a distribution with no variance.
First, we see that the mitigation approaches lead to a spead-up, that is more prominent for short lifetimes.
A reason for this can be that the difference between the approaches gets amplified due to the resets caused by memory time-outs.
Thus, our manual mitigation strategies provide an important advantage even in the presence of significant temporal losses.

We note that our brute force optimization approach suffers from the noise introduced by the Monte Carlo sampling.
Due to the presence of significant noise, the algorithm can terminate early if the mean does not change significantly.
This means that our approach cannot guarantee to find the ideal solution.
In this case but also when the lifetime becomes irrelevant because it largely exceeds the expected entanglement generation duration,
we expect our method to systematically underestimate the optimal cut-off window duration, see Fig.~\ref{fig:opt-duration-decay}(c).
This is also visible from the jump in all panels of Fig.~\ref{fig:opt-duration-decay} around a lifetime of \(10^{-0.75}\,\)s.
Considering that the jump does not significantly affect the trend of the average duration [in panel (a) of Fig.~\ref{fig:opt-duration-decay}] we still believe that our results are indicative of the achievable performance.

Finally we note that as seen in Fig.~\ref{fig:opt-duration-decay}(b), the average emission probability increases as a function of lifetime, until it saturates for long lifetimes.
This is in agreement with our expectation that as the lifetime exceeds the average duration of the whole protocol, the contribution to the infidelity from the inhomogeneity induced by the decay to the memory efficiencies becomes negligible and thus the ideal emission probabilities are given by the static system.

\subsection{Other errors}

In the preceding analysis we focused on the non-uniform memory efficiencies and thus neglected other detrimental effects.
In the regime of high target fidelities all errors have to be small, such that they may be treated perturbatively.
To first order they can thus be treated as additive (as we have used in the analytical analysis).
Furthermore, we already included dark counts in the model, making it possible to add the effect of dark counts if necessary.
Here, we only stress that the detector dark count rates need to be sufficiently low, not to limit the fidelity of the generated entangled states.

In a more complete descriptions other errors should be added,
in particular they can arise from
other detector imperfections,
random phase fluctuations making phase stabilization necessary \cite{sangouard-2011-quant_repeat_based_atomic_ensem_linear_optic,beukers-2024-remot_entan_protoc_station_qubit},
and imperfect photon indistinguishability \cite{sangouard-2011-quant_repeat_based_atomic_ensem_linear_optic,avis-2024-asymm_node_placem_fiber_based_quant_networ}
where linear optics or photon shaping techniques can be used for mitigation
\cite{vasilev-2010-singl_photon_made_to_measur,farrera-2016-gener_singl_photon_with_highl,morin-2019-deter_shapin_reshap_singl_photon,tissot-2024-effic_high_fidel_flyin_qubit_shapin,omlor-2025-entan_gener_using_singl_photon,cussenot-2025-unitin_quant_proces_nodes_cavit}.

In total, we believe that this list of further imperfection shows that our analysis and mitigation strategies should be compatible and complement efforts to improve detectors, increase phase stability, and photon indistinguishability.

\section{Conclusion}\label{sec:conclusion}

We have studied the effect of non-uniform memory efficiencies on the DLCZ protocol implemented with photonic memories.
Based on our analysis, we introduced various mitigation approaches, that can be used to achieve good rate and fidelity of entanglement generation even in the presence of strong variations in device-to-device memory efficiency.
From our mitigation approaches, we can see that the placement of nodes is a crucial resource, but also that having varying photon emission probabilities of the nodes during initial entanglement generation is a simple and valuable resource for mitigating differences in efficiencies.
Furthermore, we found that the enhancement persists when we also consider memory decay. 
Here, we even observed an amplification of the benefit of the mitigation strategies for short lived memories.
With our experimentally friendly  mitigation strategies of sorting memories and adjusting emission probabilities we are able to achieve a performance comparable to and often even better than obtained for identical memories. This is in contrast to the naive approach of random placement and uniform emission probabilities, which often failed to reach the target fidelity. This indicate that such mitigation strategies may be indispensable for achieving the desired performance in the lab.
All-in-all, we believe that our analysis and mitigation strategies, will provided a valuable tool that complements other efforts in improving entanglement generation within quantum networks.
As a full optimization in presence of decay proofs to be challenging due to the noise introduced by Monte Carlo sampling, an interesting challenge for future research is the development of noiseless alternative methods to this sampling.

We also emphasize that we only consider a specific sorting of the memories
(which we found to perform favorable for a small sample case, and can motivate from the analytical considerations).
It may still be possible to achieve a better performance by also optimizing over various placements.
Furthermore, we have here only considered a single DLCZ-protocol and a short repeater chain with only a single repeater station.
For communication over larger distances, it would be interesting to investigate longer repeater chains as well as the many modifications of the DLCZ-protocol.

\begin{acknowledgments}
  We thank E. R. Hellebek, S. Maiti, M. van Hooft, and C. Gustin for insightful discussions.

  This work was funded by the European Union's Horizon Europe research and innovation programme under grant agreement No. 101102140 – QIA Phase 1.
  Funded by the European Union. Views and opinions expressed are however those of the author(s) only and do not necessarily reflect those of the European Union or European Commission. Neither the European Union nor the granting authority can be held responsible for them.
  
  We acknowledge the support of Danmarks Grundforskningsfond (DNRF Grant No.~139, Hy-Q Center for Hybrid Quantum Networks).
\end{acknowledgments}

\appendix

\section{Emission Probability Optimization and Simulation}\label{app:opt}
In this appendix we provide additional details on the implementation of the manual optimization based on pre-determined relations between the emission probabilities as well as on the full numerical optimization for static memory efficiencies.

\subsection{Emission probability optimization for heuristic approaches}
For the heuristic approaches \ref{list:uniform}, \ref{list:symmetric}, and \ref{list:biased} we choose a fixed relation between the emission probabilities and memory efficiencies of the different sites, see Sec.~\ref{sec:mitigation}, and
we can use a tailored optimization approach.
The idea behind this approach is that we assume that if the emission probabilities have a fixed ratio, the fidelity of the whole system can be expressed in terms of
\(F(p) = F_0 + a(p) p\), with a constant part \(F_0\) and a part that is monotonically decreasing \(a(p) < 0\)  when increasing the average emission probability \(p\).
Furthermore, we assume that we target a high efficiency \(0 < 1 - F_{\text{target}} \ll 1\) and that \(a(p)\) varies slowly with \(p\), i.e., \(\frac{d a(p)}{d p} \ll a(0)\).
Note, that the former condition also necessitates \(0 < 1 - F_0 \ll 1\).

With these assumptions we can approximate the iterative steps in Newton method's \cite{press-2007-numerical_recipes} to find the root of \(F(p) - F_{\text{target}}\) as
\begin{align}
  \label{eq:manual_opt_p}
  p_{\text{i+1}} = & \frac{1 - F_{\text{target}}}{1 - F(p_i)} p_i X ,
\end{align}
where in the numerical implementation we use \(X = 1\).
Higher order correction factors in the non-linearity of the fidelity and the intrinsic infidelity can be accounted for by using
\( X = \left[ 1 - \frac{a'(p_i)}{a(p_i)} \right]  \left\{ 1 + \frac{F_{\text{target}} - F(p_i)}{[1-F(p_i)] (1 - F_{\text{target}})} (1-F_0) \right\} \).
We terminate the optimization if \(0 \le F(p_i)/F_{\text{target}} - 1 < \epsilon\), and choose \(\epsilon = 10^{-4}\).

\subsection{Numerical Optimization}
For the full numerical parameter optimization we make use of the ``NelderMead()'' method implemented in ``Optim.jl'' \cite{mogensen-2018-julia_optim} in the Julia programming language \cite{bezanson-2017-julia},
where the cost function is based on the duration, but we also add a cost to account for constraints.
Constraints (e.g., the fidelity bound) can be violated in this approach, but this is suppressed by the additional cost.
We use an exponential cost for violating the target fidelity, and a linear cost for violating the allowed emission probabilities which we restrict to be in the interval \((0,0.1]\).
The full implemetation is available as Ref.~\cite{CODE}.

\begin{figure*}
\centering
\includegraphics[width=\linewidth]{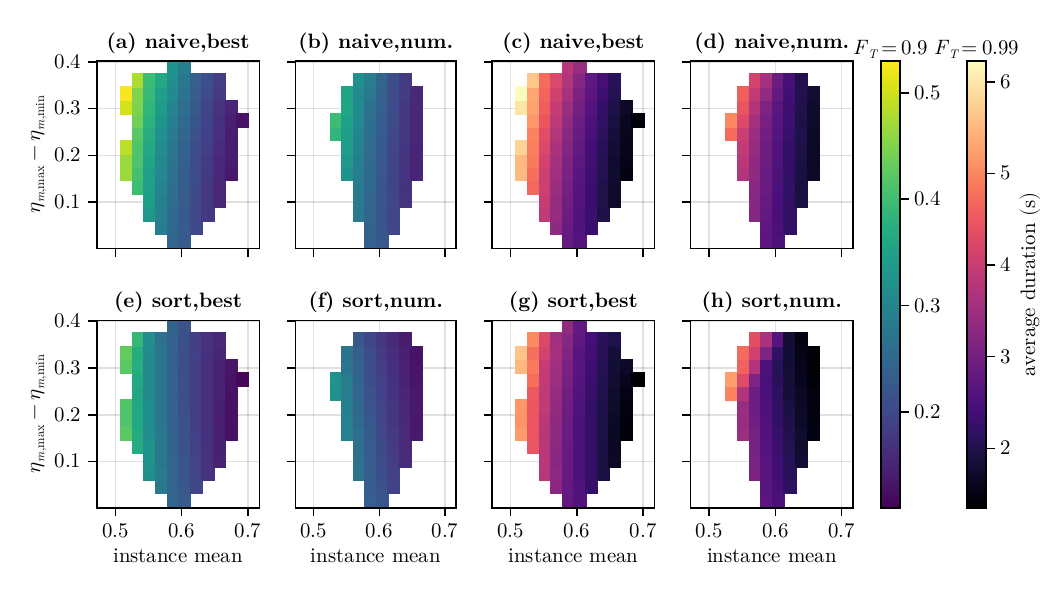}
\caption{\label{fig:var-mean-dur} Average duration as a function of memory variance and mean (of the instance).
  We bin the data of Fig.~\ref{fig:opt-duration}, according to instance mean and instance variance quantified by \(\eta_{m,\max} - \eta_{m,\min}\).
  In the first column (a-d) we display results for random memory placement, while in the second column (e-h) we display results following the sorted placement approach \ref{list:sort}.
  The even rows (a,e,c,g) display the data according to the best heuristic approach, while the odd rows (b,f,d,h) display the data for the full numeric optimization.
  The four left panels (a,b,e,f) show the results for a target fidelity $F_T = 0.9$ and the right panels (c,d,g,h) for $F_T = 0.99$.
}
\end{figure*}

\section{Average duration binned according to instance memory efficiencies}\label{app:binning}
In the main text we focused on the results averaged over samples drawn from a fixed distribution as functions of the parameters determining the distribution.
This implies, that for the different datapoints, instances with different instance-mean memory efficiency
and different instance-variance contribute.
To isolate these effect, we here bin the data according to the values of each instance (apart from the 0 variance points).
For this binned approach we display the average duration according to both the best semi-analytic and full numerical duration in Fig.~\ref{fig:var-mean-dur}.

We bin the data into 15 bins for both the difference \(\eta_{m,\max} - \eta_{m,\min}\) between the largest and lowest memory efficiency of the instance, and the mean of the instance.
To ensure a degree of statistical significance, we only display bins that have more than \(50\) data points.

In the data we see a clear effect of the duration being shorter if the particular samples have a larger average efficiency, i.e., if we are lucky to draw memories with high efficiency.
There is only a minor dependency on the variance (and only for the numerically optimized approaches) which tend to prefer similar memory efficiencies.
Overall, these results thus support the conclusion reached in the main text, that the various mitigation strategies are largely able to counteract the effect of differences in memory efficiencies and reach a similar performance as systems with homogeneous memory efficiencies.
In the main text, we therefore plot the results according to how they were obtained in the numerical simulation, although this means that we bin data with different instance mean, which is the main parameter affecting the duration.

\bibliography{refs.bib}

\end{document}